\documentclass[10pt]{article}
\usepackage[OE]{express}
\usepackage{cite}
\usepackage{amsmath,amssymb}
\usepackage[latin1]{inputenc} 
\usepackage{dsfont}
\makeatletter
\def\blfootnote{\xdef\@thefnmark{}\@footnotetext}
\makeatother

\begin{document}
\title{Performance analysis of FSO communications under LOS blockage}

\author{Jose Maria Garrido-Balsells,\authormark{1} F. Javier Lopez-Martinez,\authormark{1} Miguel Castillo-Vazquez,\authormark{1} Antonio Jurado-Navas,\authormark{1} and Antonio Puerta-Notario\authormark{1}}

\address{\authormark{1}Dpt. Communications Engineering, University of M\'alaga, Campus Teatinos s/n, E-29071 M\'alaga, Spain}

\email{\authormark{*}jmgb@ic.uma.es} 



\begin{abstract}
We analyze the performance of a free-space optical (FSO) link affected by atmospheric turbulence and line-of-sight (LOS) blockage. For this purpose, the atmospheric turbulence induced fading is modeled by the $\cal M$-distribution, which includes the Gamma-Gamma distribution as special case. We exploit the fact that the physical interpretation of the $\cal M$-distribution allows to split the optical energy through the propagation link into three different components: two coherent components and one incoherent scatter component. Based on this separation, we derive novel analytical expressions for the probability density function (PDF), for the cumulative distribution function (CDF) and for the moment generating function (MGF) of the $\cal M$-distribution under the temporary blockage of the coherent components, hereinafter referred to as LOS blockage. Further, a new closed-form expression for the outage probability (OP) under LOS blockage is derived in terms of the turbulence model parameters and the LOS blockage probability. By means of an asymptotic analysis, this expression is simplified in the high-SNR regime and the OP in terms of the diversity order and diversity gain is then deduced. Obtained results show that the impact of the LOS blockage on the OP strongly depends on the intensity of the turbulence and on the LOS blockage probability.\blfootnote{ This work has been submitted to for journal publication. Copyright may be transferred without notice, after which this version may no longer be accesible}
\end{abstract}




\section{Introduction}

Free-space optical (FSO) communications are currently a solid alternative to radio communications in many applications including last mile broadband access, cellular backhaul, inter-building connections and high-speed links of next generation all-optical networks \cite{Kaz10}. The main advantage of FSO systems lies in the possibility of exploiting the huge unregulated bandwidth available in optical frequencies that allows data rates similar to those of fiber optic systems \cite{Ciaramella}.

However, in FSO links the transmitted signal is affected by various factors before arriving at the receiver that degrade the link capacity and limit the transmission rate. These include atmospheric loss, atmospheric turbulence induced fading, and blockage. 
First, the high attenuation introduced by fog, rain or pollution in FSO communications causes that the practical distance between transmitter and receiver is limited up to several kilometers. A detailed study of atmospheric loss can be found in \cite{Bloom}. Nevertheless, the most impairing atmospheric effect is produced by random changes of the medium refractive index along the propagation path. This effect known as atmospheric turbulence leads to the appearance of random fading intervals in the received optical irradiance. 

Extensive research has been performed by the scientific community over the years in the search of a statistical distribution capable to model these irradiance fluctuations under any type of turbulence conditions. As a result of this research, different mathematical models for the probability density function (PDF) of the received optical irradiance have been proposed so far \cite{And05,Zhu02,AlH01,Churnside, Toyo11, CapLibro}. Such models adapt to a particular range of turbulence intensities, i.e. the Log-normal \cite{Zhu02} distribution was proposed to model weak turbulence conditions, whereas the Gamma-Gamma \cite{AlH01} model was proposed to describe the irradiance behavior under moderate to strong turbulence conditions. Due to their mathematical tractability, these two distributions have been widely used to model the turbulence induced fading of the FSO links. More recently, an alternative statistical distribution, called $\cal M$ or Malaga, was proposed in \cite{CapLibro}. This distribution unifies in a closed-form expression most of the statistical models for the irradiance fluctuations proposed in the literature, including the popular Gamma-Gamma distribution. The $\cal M$-distribution is applicable to unbounded optical wavefront transmission under the full range of atmospheric turbulence intensities, as detailed in \cite{CapLibro}. 
The main characteristic of this new model is the definition of three different optical components in the signal traveling through the atmosphere: a line-of-sight (LOS) component and two scattered components. The first scatter component is coupled to the LOS component whilst the second one is the classical independent non-LOS (NLOS) scattering component. In \cite{OE_MKg}, a reformulation of the $\cal M$-distribution was presented and new and simpler analytical expressions for the model were proposed based on a mixture of the known Generalized-$K$ and discrete Binomial and Negative Binomial distributions. Note that the Generalized-$K$ distribution was introduced by Barakat \cite{Barakat86} and Jakerman \cite{Jakerman87} to describe atmospheric laser transmissions under weak scattering conditions, and also applied in radar applications \cite{Shankar04, Bithas06}. 
 
In addition to the aforementioned atmospheric factors, FSO links are also sensitive to blockage due to moving objects intersecting the laser beam propagation path \cite{Ghassemlooy2009}. These interruptions are quite frequent \cite{Kolka} and, therefore, their effect cannot be neglected in practical systems.  
For this reason, the performance analysis of FSO links with link blockage was recently addressed in \cite{Djordjevic16}. Here, assuming a Gamma-Gamma fading model for the irradiance fluctuations, it was concluded that the link blockage causes the appearance of an outage floor in the high-SNR regime which is coincident with the blockage probability. This means that the OP in the presence of blockage cannot be improved by increasing the transmission power. However, the Gamma-Gamma model used in \cite{Djordjevic16} is not the most appropriate to analyze the effect of blockages since it does not allow to accurately consider real scenarios in which only a partial obstruction of the laser beam occurs.



To cover these cases, in this paper, we extend the analysis presented in \cite{Djordjevic16} and examine the effect of partial blockage in the performance of FSO links affected by turbulence fading. Unlike in \cite{Djordjevic16}, the $\cal M$-distribution is assumed here to model the atmospheric turbulence induced fading. Thus, by using this distribution, we can explicitly consider the partial blockage of the laser beam. In particular, we analyze a type of partial blockage in which the coherent component of the laser beam is obstructed. This case corresponds to the possibility of a small\footnote{We refer to small objects compared to the divergence of the laser beam, which can block the transverse coherence radius but not the received power through NLOS scattering.} object temporarily obstructing the propagating path within the transverse coherence radius \cite{And05}. Hereinafter, we refer to this type of partial blockage as LOS blockage. As a result of this analysis, a novel closed-form expression for the outage probability in terms of the turbulence model parameters and the LOS blockage probability is derived. Further, 
when LOS blockage occurs, the OP does not reach an irreducible error in the high-SNR regime. Thus, we can calculate the power offset required to overcome the effect of blockage in a very simple form. 
Note that since the Gamma-Gamma distribution is a particular case of the $\cal M$-distribution, the results derived here reduce to those presented in \cite{Djordjevic16} when the coupling parameter of the $\cal M$-distribution is equal to one. 

The remainder of this paper is organized as follows. Section 2 describes the system model and the different types of link blockage. In section 3, the PDF and MGF of the fading model under LOS blockage is derived. Section 4 focuses on the link outage analysis. Finally, results and most important conclusions are presented in section 5. 

\section{System model and LOS blockage}\label{Sec2}
In this work, we assume an on-off keying (OOK) intensity modulation with a direct detection (IM/DD) scheme, so that the received optical irradiance, $I_{rx}$, is the variable to be considered. Thus, the instant photocurrent at the detector output is given by
\begin{equation}
y(t)= R\cdot I_{rx}(t) + n(t)
\end{equation}
\noindent where $R$ is the detector responsivity, which here is assumed equal to 1, and $n(t)$ is the noise at the receiver, which is modeled as additive white Gaussian noise (AWGN) with zero mean and variance $\sigma_n^2$ \cite{Hranilovic}. 
For a point-to-point optical communication link, due to the multiplicative effect of the random fading induced by the atmospheric turbulence \cite{And05}, the value of  $I_{rx}$ can be obtained from the product of the received irradiance in absence of atmospheric turbulence, $I_0$, and the normalized optical irradiance, $I$,  i.e. $I_{rx}=I_0 I$. Note that here, $I_0$ is a deterministic value which includes the atmospheric loss while $I$ is a random variable (RV) with statistical average $\text{E}[I]= 1$ which includes the refractive and diffractive effects of the atmospheric turbulence cells. Then, the electrical signal-to-noise ratio (SNR) at the receiver can be expressed as 
\begin{equation}\label{EqSNR}
 \gamma=\frac{(RI_0I)^2}{\sigma_n^ 2}=\gamma_0 I^ 2
\end{equation}
\noindent where the parameter $\gamma_0$ represents the received electrical SNR in the absence of atmospheric turbulence.

\subsection{Turbulence fading model}
\begin{figure}[t]
    \centering
           \includegraphics[width=10cm]{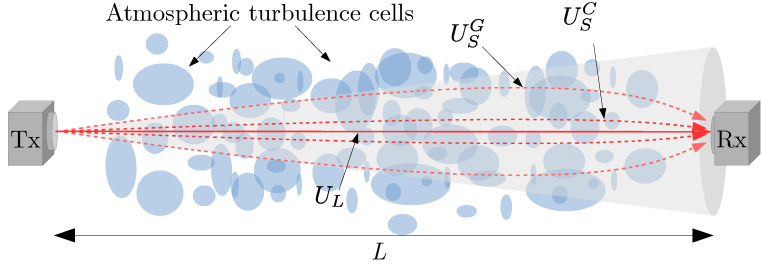}
    \caption{${\cal M}$-model laser beam propagation scheme. The three components received are: the line-of-sight (LOS) term, $U_L$, the coupled-to-LOS scattering term, $U_S^C$ and the classic scattering term related to the off-axis eddies, $U_S^G$.}
    \label{FigModelo}
\end{figure}

In order to model the optical irradiance fluctuations caused by the atmospheric turbulence, the statistical $\cal M$-distribution is here assumed. This distribution relies on the use of doubly stochastic theory of scintillation in which the large- and small-scale turbulence eddies are supposed to induce refractive and diffractive effects on the light beam \cite{And05}. Thus, the normalized received irradiance $I$ is considered as the product of two independent random variables $X$ and $Y$, which represent the irradiance fluctuations arising from large- and small-scale fluctuations, respectively, as follows
\begin{equation}\label{Eq_I}
I=Y\cdot X=\left|U_L+U_S^C+U_S^G \right|^2 \exp\left(2\chi\right),
\end{equation}  
\noindent 
where $X=\exp(2\chi)$ and $Y=\left|U_L+U_S^C+U_S^G \right|^2$. As detailed in \cite{CapLibro,OE_MKg}, the RV $X$ follows a Gamma distribution, while $Y$ has the form of a Rician-Shadowed distribution.
In this scheme, the small-scale fading characteristic of the atmospheric channel is modeled by three different signal components: a main LOS component, denoted as $U_L$, and two additional scattering components $U_S^C$, and $U_S^G$, as  shown in Fig. \ref{FigModelo}.
The first scattering component, $U_S^C$, is coupled to the LOS component and it is generated by the on-axis cells, i.e. by the quasi-forward eddies on the propagation axis. This fact makes this scattered optical field to coherently contribute to the LOS term. 
The second scattering component, $U_S^G$, is the classical NLOS optical signal due to the energy scattered by the off-axis eddies which is statistically independent of the other two components. 

The $\cal M$-distribution is characterized by 3 parameters: $\alpha$, $\beta$ and $\rho$. The parameter $\alpha$ is related to the effective number of large-scale eddies, the parameter $\beta$ is related to the amount of fading (AF) introduced by the diffraction effects associated to the small-scale eddies, and the coupling parameter $\rho$ represents the amount of scattering power coupled to the LOS component and ranges from 0 to 1, i.e. $0\leq \rho \leq 1$. Note that for $\rho\rightarrow1$ the $\cal M$-distribution reduces to the Gamma-Gamma distribution with parameters $\alpha$ and $\beta$. 

Following the notation of \cite{OE_MKg}, the average optical power of the LOS component is given by  $\Omega=\text{E}[|U_L|^2]$, while the average power contained in both scattering components is ${\xi=\text{E}[|U_S^C|^2+|U_S^G|^2]}$. In turn, the power of the coupled-to-LOS and classic scattering components can be expressed as a function of the coupling parameter, $\rho$, as $\xi_c=\rho \xi$ and $\xi_g=(1-\rho) \xi$, respectively, being $\xi=\xi_c+\xi_g$. Finally, the average optical power associated to the coherent contribution of the LOS and the coupled-to-LOS components is denoted as $\Omega'$. It is worth noting that although the received optical field is described here as the summation of three signals, it can also be physically interpreted as a summation of two effective optical components: the coherent component, composed of LOS and coupled-to-LOS terms, with average power $\Omega'$, and the incoherent component, with average power $\xi_g$. This alternative description will be used in the next section. Note that since we assume that the total average optical irradiance received is normalized, then $\text{E}[I]=\Omega'+\xi_g=1$. 

According to the reformulation of the $\cal M$-distribution \cite{OE_MKg}, the PDF of the normalized irradiance $I$ can be written in a compact form as a mixture of the Generalized-$K$ distribution with a Binomial or Negative binomial discrete distribution as follows
\begin{equation}\label{EqMPDF}
f_I(I)=\sum\limits_{k = 1}^{\widetilde{k}}  {\widetilde{m}_k K_G(I;\alpha,k,\widetilde{\mu}_k)},
\end{equation}
\noindent where the set of parameters $\{\widetilde{k}, \widetilde{m}_k, \widetilde{\mu}_k\}$ depends on the $\cal M$-distribution parameters $\{\alpha, \beta, \rho\}$. 
As detailed in  \cite{OE_MKg}, Eq.\eqref{EqMPDF} can be seen as
the superposition of $\widetilde{k}$ Generalized-$K$ sub-channels, $K_G(I;\alpha,k,\widetilde{\mu}_k)$, corresponding each one to a different physical optical path. Every sub-channel is weighted by a Binomial distribution $\widetilde{m}_k$ which depends on the inherent parameters of the $\cal M$-distribution and is given by 
\begin{equation}\label{Eqmk_p}
\widetilde{m}_k=\begin{cases}
{\binom{\beta  - 1}{k - 1}} p^{k-1}(1-p)^{\beta-k} & \beta\in \mathbb{N} \\ 
\frac{\Gamma(k-1+\beta)}{\Gamma(k)\Gamma(\beta)}p^{k-1}(1-p)^\beta & \beta\in \mathbb{R}
\end{cases},
\end{equation}
\noindent where the parameter $p$ can be interpreted as the success probability in a Bernoulli experiment in which a success event is considered when the signal travels through the on-axis coherent components, and a failure event is defined when the signal takes part of the off-axis independent term. This parameter is obtained as
\begin{equation}\label{Eqp}
p=\left[{1+\left(\frac{1}{\beta}\frac{\Omega'}{\xi_g}\right)^{-1}}\right]^{-1}.
\end{equation}

The value of $\widetilde{k}=\beta$ when $\beta \in \mathbb{N}$ and $\widetilde{k}\to \infty$ when $\beta \in \mathbb{R}$. However, as detailed in \cite{OE_MKg}, it is possible to find a $\widetilde{k}=k_{max}$ which assures that the difference between the truncated and the exact formulation is bounded by an specified error tolerance parameter $\epsilon$, i.e. for $k\geq k_{max}$ the cumulative distribution function (CDF) of the corresponding $\widetilde{m}_k$ distribution is higher than $(1-\epsilon)$. Note also that, as explained in \cite{OE_MKg}, the lower sub-channel orders are associated to more adverse turbulence conditions.

The average optical irradiance for each Generalized-$K$ sub-channel, $\widetilde{\mu}_k$, is expressed as
\begin{equation}\label{Eqmu_Kg}
\widetilde{\mu}_k=\begin{cases}
\frac{k}{\beta}\left(\xi_g\beta+\Omega' \right) & \beta\in \mathbb{N} \\ 
k\xi_g & \beta\in \mathbb{R} 
\end{cases},
\end{equation}
\noindent and, finally, the PDF of the irradiance traveling through the $k$-th sub-channel is given by the generalized-$K$ distribution as
\begin{equation}\label{EqPDFKg}
K_G(I;\alpha,k,{\cal I}_k)=\frac{2B^{(\alpha+k)/2}}{\Gamma(\alpha)\Gamma(k)}I^{(\alpha+k)/2-1}K_{\alpha-k}\left(2\sqrt{B\cdot I}\right)
\end{equation}
\noindent where $B=\alpha k/{\cal I}_k$, with ${\cal I}_k=\text{E}[I]$ being the average optical irradiance through the $k$-th sub-channel, and $K_{\nu}(\cdot)$ is the modified Bessel function of the second kind. It must be noted that the generalized-$K$ distribution is characterized by two shape parameters, i.e. $\alpha$ and $k$, that can be modified to describe different fading and shadowing scenarios \cite{Shankar04}. As verified in \cite{OE_MKg} the higher the sub-channel order $k$ is, the lower the amount of fading parameter (AF), thus implying that those sub-channels with a higher order are better for transmission. Conversely, a lower order is related to those sub-channels with worse atmospheric conditions and thus, with higher AF values. 

\subsection{LOS blockage}
\begin{figure}[t]\label{FigBlockageModel}
	\centering
	\includegraphics[width=8.5cm]{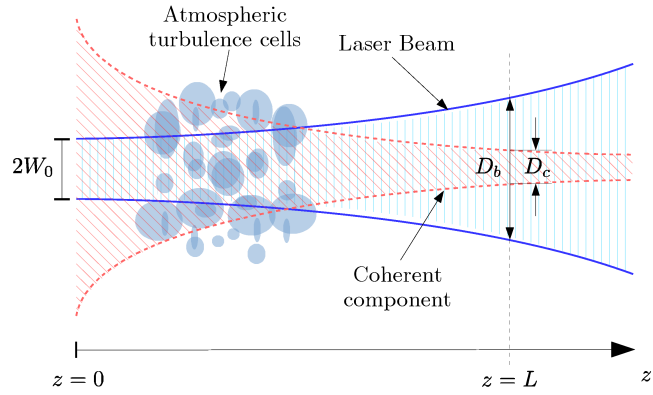}
	\vspace{0.1cm}\\
	(a)\\
	\vspace{0.1cm}
	\includegraphics[width=9cm]{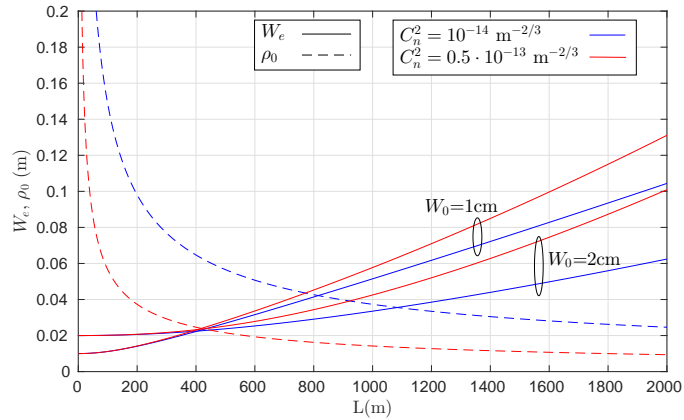}
	\vspace{0.1cm}\\
	(b)\\
	\vspace{0.1cm}
	\caption{(a) Propagation model of a partially coherent Gaussian laser beam: $W_0$ is the initial beam diameter, $D_b=2W_e$ is the beam diameter and $D_c=2\rho_0$ is the transverse coherence diameter, both at $z=L$. (b) Laser beam radius, $W_e$, and transverse coherence radius, $\rho_0$, as a function of the propagation length, $L$, for moderate and strong turbulence conditions.}
\end{figure}

In an optical link, the width of the light beam produced by a transmitter laser expands as the propagation distance increases, as shown schematically in Fig. \ref{FigBlockageModel}(a) in blue color.  For horizontal FSO transmissions, a good approximation is to consider a Gaussian profile for the beam intensity \cite{Majumdar}, so that in the absence of atmospheric turbulence, the radius of this beam for a distance $L$ is given by \cite{And05}
\begin{equation}
\label{Beam}
W(L)=W_0\sqrt{\left(1-\frac{L}{F_0}\right)^2+\left(\frac{2L}{\text{k}W_0^2}\right)^2},
\end{equation}
\noindent 
where $W_0$ is the beam radius  at the transmitter, i.e. at $z=0$, $F_0$ is the phase front radius of curvature at the transmitter output, $\text{k}=2\pi/\lambda$ is the wave number and $\lambda$ is the wavelength. From Eq.\eqref{Beam}, it follows that, during the designing of a FSO link, it is possible to control the beamwidth produced at a certain distance by adjusting properly the laser parameters $W_0$ and $F_0$. For example, assuming a collimated beam (i.e. $F_0\to \infty$), laser divergence can be reduced increasing the value of the initial radius $W_0$ by means of a beam expander. 

However, in the presence of atmospheric turbulence,  Eq.\eqref{Beam} is no longer valid and the beam spreading increases even more. In this case, a new effective beam radius $W_e$, greater than $W$, is defined \cite{And05} as
\begin{equation}
\label{BeamTurbulent}
W_e(L)=W(L)\sqrt{1+1.625 \sigma_1^{12/5} \Lambda},
\end{equation}
\noindent
where $W(L)$ is the beam radius in the absence of turbulence at $z=L$, given by  Eq.(\ref{Beam}), $\sigma_1^2$ is the Rytov variance and $\Lambda= 2L/\text{k}W^2$. Note that the value of the Rytov variance is obtained with the expression $\sigma_1^2=1.23C_n^2\text{k}^{7/6}L^{11/6}$, being $C_n^2$ the refraction index structure parameter, which is directly related to the turbulence strength \cite{And05}.

In addition to the increase in the laser beam width, the atmospheric turbulence also destroys the spatial coherence of a laser beam as it propagates through the atmosphere. In particular, the transversal coherence radius $\rho_0$, which defines the coherent component of the beam, is reduced as the distance and the turbulence strength increase. This reduction is schematically depicted in Fig. \ref{FigBlockageModel}(a) where the coherent component of the beam is represented in red. As reported in \cite{And05}, assuming a plane wave (i.e. far from the transmitter), the spatial coherence radius at a distance $L$ is given by
\begin{equation}
\label{Ro}
\rho_0(L)=\left(1.46C_n^2\text{k}^2L\right)^{-3/5} 
\end{equation}
\noindent where $l_0<<\rho_0<<L_0$, and $l_0$ and $L_0$ are the inner and outer scales of the turbulence, respectively.

Consider now an obstacle of diameter $D$ between the transmitter and the receiver. Depending on the size of this obstacle compared to the laser beam diameter $D_b=2W_e$ and to the coherent beam component diameter $D_c=2\rho_0$, two types of blockages can be occur. On the one hand, when $D \ge D_b$, a total link blockage is produced and no signal is detected at the receiver. The effect of this type of total blockage on the FSO link performance is examined in \cite{Djordjevic16}. On the other hand, when $D<D_b$, a partial link blockage is produced and, unlike in the previous case, one part of the signal still arrives at the receiver through the non-coherent component. In the case of partial blockage in which $D=D_c$, only the coherent component of the laser beam is blocked and, as mentioned above, we refer to this type of partial blockage to as LOS blockage.
In order to illustrate the evolution of $W_e$ and $\rho_0$ with the propagating length and to obtain realistic values of $D_b$ and $D_c$, we have depicted in Fig. \ref{FigBlockageModel}(b) the values of Eqs. (\ref{BeamTurbulent}) and (\ref{Ro}) for moderate ($C_n^2=10^{-14}$) and strong ($C_n^2=0.5\cdot 10^{-13}$) turbulence intensities, assuming a wavelength $\lambda=1550$ nm. In addition, two different transmitter beam radius have been applied ($W_0=1$ cm and $W_0=2$  cm). From this figure, it is observed that for moderate turbulence conditions and a laser with $W_0=1$ cm, at a distance $L=1600$ m, an obstacle with $D=16$ cm will produce a total blockage. Likewise, a smaller obstacle with $D=6$ cm will cause a LOS blockage. For strong turbulence, at a half distance ($L=800$ m), the obstacle diameters would be of $D=9$ cm and $D=3$ cm for total and LOS blockage cases, respectively. From these results, it should be noted that both total and LOS blockage are feasible considering real obstruction sizes such as the ones produced by small birds.

Finally, since the spatial coherence diameter, $D_c$ decreases as distance increases, the probability of occurrence of LOS blockage, $P_b$, grows for large propagation lengths This probability is defined as the obstruction probability of the transverse coherence area along the propagation path. The value of $P_b$ depends on different factors such as the laser parameters, birds density and size, vegetation, etc. Some practical values of the probability of total blockage based on measurements are presented in \cite{Kolka}. Throughout this paper, we will analyze in depth the effect of LOS blockage in FSO links for a given blockage probability, $P_b$. 

\section{PDF and MGF of the turbulence fading model under LOS blockage}
In this section, we derive de PDF and the MGF of the $\cal M$-distribution considering the effect of the LOS blockage. These functions will be used in next section to perform an outage analysis.

\subsection{PDF of the $\cal M$-distribution under LOS blockage}
Here, we modify the turbulence fading model described in section 2 in order to include the effect of the LOS blockage. Then, the PDF of the combined model with both effects the turbulence and LOS blockage is derived. It is worth noting that the inclusion of the LOS blockage on the $\cal M$-distribution is possible due to its inherent physical interpretation, reported in \cite{CapLibro}. According to this interpretation, we can assume that LOS ($U_L$) and coupled-to-LOS ($U_S^C$) optical terms contain the energy coming from the coherent component of the laser beam and, similarly, we can also assume that the scattered term ($U_S^G$) contains the optical energy coming from the area outside the coherent component of the laser beam. Thus, under these assumptions, the LOS blockage effect can be easily included in the $\cal M$-distribution by adding a discrete RV $b$ in Eq.(\ref{Eq_I}) as follows:
 \begin{equation}\label{Eq_Imod}
I=\left|b \left(U_L+U_S^C\right)+U_S^G \right|^2 \exp\left(2\chi\right).
\end{equation}  
\noindent
Note that the RV $b$ can take two possible values, $b=0$ or $b=1$, depending on whether a blockage of LOS components is considered or not, respectively. As previously discussed, the blockage probability (i.e. the probability of $b=0$) depends on the specific configuration of the optical link.

With this change in Eq.\eqref{Eq_I}, when a LOS blockage is produced, the combined power of the LOS and couple-to-LOS terms is forced to zero, i.e. $\Omega'=0$, and then the normalized received irradiance, $I$, will only contain the power of the independent scattered term $\xi_g$. As a consequence, the Eqs. \eqref{Eqmk_p}, \eqref{Eqp}, and \eqref{Eqmu_Kg} which define the weight coefficients $\widetilde{m}_k$,  the $p$ parameter and the average optical powers $\widetilde{\mu}_k$, respectively, are now simplified. More precisely, the parameter $p$, which is related to the probability of the optical signal to travel coupled to the LOS term within the coherence transverse disc, is now equal to zero, since
\begin{equation}\label{Eq_p_blockage}
p=\left[{1+\left(\frac{1}{\beta}\frac{\Omega'}{\xi_g}\right)^{-1}}\right]^{-1}=\frac{\Omega'}{\Omega'+\beta \xi_g}=0.
\end{equation}
Further, after substituting $p=0$ into Eq. \eqref{Eqmk_p}, the mixture weight coefficients given by $\widetilde{m}_k$ take now the values
\begin{equation}\label{Eqmu_Kg2}
\widetilde{m}_k=\begin{cases}
1 & k=1\\ 
0 & k\neq 1 
\end{cases},
\end{equation} 
\noindent what indicates that, regardless of the value of $\beta$, the only active Generalized-$K$ sub-channel is the first-order one, in which the transmitted optical power is $\widetilde{\mu}_1=\xi_g$. Note that this result is in perfect agreement with the blockage scenario here described, and with the Generalized-$K$ distribution behavior, which models a non-LOS Rayleigh fading channel when $k=1$ \cite{OE_MKg, Shankar04}. Thus, considering all these simplifications, when a LOS blockage is produced, then $f_I(I)=K_G(I;\alpha,1,\xi_g)$.

Therefore, if a non-zero $P_b$ is assumed, the modified PDF of the normalized received optical irradiance $I$ can be expressed as
\begin{equation}\label{Eq_MpdfPb}
f_{I,b}(I)=P_bf_I(I|b=0)+(1-P_b)f_I(I|b=1),
\end{equation}
\noindent where $f_I(I|b=0)$ stands for the PDF of the received irradiance under blockage and $f_I(I|b=1)$ stands for the general case of no blockage, given by the PDF of Eq. \eqref{EqMPDF}. Substituting the values of both functions, we obtain the PDF of the $\cal M$-distribution under LOS blockage as
\begin{equation}\label{Eq_MpdfBloc}
f_{I,b}(I)=P_b K_G(I;\alpha,1,\xi_g) +(1-P_b) \sum\limits_{k = 1}^{\widetilde{k}}  {\widetilde{m}_k K_G(I;\alpha,k,\widetilde{\mu}_k)}.
\end{equation} 
From Eq. \eqref{Eq_MpdfBloc} it follows that the independent scattered component power, $\xi_g$, is transmitted through the first-order sub-channel when LOS blockage is considered, whereas under a non-blockage scenario this power is split among all the active Generalized-$K$ sub-channels, with probabilities controlled by the discrete mixture coefficients $\widetilde{m}_k$.

It should be pointed out here that possible contributions to the independent scattered signal term produced by the refractive and diffractive phenomena induced by the obstacle \cite{Lerner70,Abbas05,Shapiro77} are not considered in Eq. \eqref{Eq_MpdfBloc}. As a consequence, we assume that the independent scattered term $U_S^G$ is not increased due to the blockage. Note that this can be regarded as a worst case scenario, since any additional contribution to $U_S^G$ component would provide higher received power and better link performances.
Finally, we also should remark that for the sake of clarity, we have not either considered misalignment effects between the transmitter and the receiver. However, these effects could be included by using the $\cal M$ distribution with pointing errors reported in \cite{OE_PointM, Ansari2016}.


\subsection{MGF of the $\cal M$-distribution under LOS blockage}
According to the definition of the moment generating function given by \cite{Bithas06}, the MGF of the  $\cal M$-distribution can be written as
\begin{equation}\label{Eq_MGF}
M_{I}(s)\triangleq \int_0^\infty \exp(-sI)f_I(I) dI,
\end{equation}
\noindent
where $f_I(I)$ is the PDF given by Eq.\eqref{EqMPDF}.  This function can be also expressed by substituting Eq.\eqref{EqMPDF} into Eq.\eqref{Eq_MGF} as follows
\begin{equation}\label{Eq_MGFM}
M_I(s)=\sum\limits_{k = 1}^{\widetilde{k}}  {\widetilde{m}_k M_{I,Kg}^{(k)}(s)},
\end{equation}
\noindent where $M_{I,Kg}^{(k)}(s)$ represents the MGF of the $k$-th sub-channel Generalized-$K$ distribution which can be calculated in the same way as in \cite[Eq. (4)]{Bithas06} as  
\begin{equation}
M_{I,Kg}^{(k)}(s)=\left(\frac{\alpha k}{\widetilde{\mu_k}s}\right)^{\frac{\alpha+k-1}{2}}\exp \left(\frac{\alpha k}{2\widetilde{\mu_k}s}\right) W_{-\frac{\alpha+k-1}{2},\frac{\alpha-k}{2}}\left(\frac{\alpha k}{\widetilde{\mu_k}s}\right),
\end{equation}
\noindent
being $W_{v,\mu}(\cdot)$ the Whittaker function \cite[Eq. (9.220)]{Gra00}, which is defined through the Tricomi confluent hypergeometric function \cite[Eq. (07.45.02.0001.01)]{Wolfram}, $U(a,b,z)$, as 
\begin{equation}
W_{v,\mu}(z)=z^{\mu+\frac{1}{2}} \exp\left(-\frac{z}{2}\right) U\left(\mu-v+\frac{1}{2}, 2\mu+1,z\right).
\end{equation}
\noindent Then, after some analytical manipulation, the MGF of the Generalized-$K$ distribution can be written as
\begin{equation}
M_{I,Kg}^{(k)}(s)=\left(\frac{\alpha k}{\widetilde{\mu_k}s}\right)^\alpha U\left(\alpha, \alpha-k+1, \frac{\alpha k}{\widetilde{\mu_k}s}\right).
\end{equation}
\noindent Note that, by definition, the use of the Tricomi confluent hypergeometric function forces the second parameter not to be integer, which means that the large-scale parameter $\alpha$ must be assumed real. This assumption is applied here to follow a convenient analytical study but the results obtained are also valid and applicable to scenarios with $\alpha \in \mathbb{N}$. In this sense, applying the primary definition of $U(a,b,z)$ given in \cite[Eq. (07.33.02.0001.01)]{Wolfram},
\begin{equation}\label{EqMGFKg}
\begin{split}
M_{I,Kg}^{(k)}(s)=&\frac{\Gamma(k-\alpha)}{\Gamma(k)}\left(\frac{\alpha k}{\widetilde{\mu_k}s}\right)^\alpha {_1}F_1\left(\alpha, \alpha-k+1, \frac{\alpha k}{\widetilde{\mu_k}s}\right) + \\ & +\frac{\Gamma(\alpha-k)}{\Gamma(\alpha)}\left(\frac{\alpha k}{\widetilde{\mu_k}s}\right)^k {_1}F_1\left(k, k-\alpha+1, \frac{\alpha k}{\widetilde{\mu_k}s}\right).
\end{split}
\end{equation} where ${_1}F_1 (�)$ is the Kummer confluent hypergeometric function of the first kind.

Finally, the MGF of the $\cal M$-distribution with blockage, given in Eq. \eqref{Eq_MpdfBloc}, can be calculated as
\begin{equation}
M_{I,b}(s)=P_b M_{I}(s|b=0)+(1-P_b)M_{I}(s|b=1),
\end{equation}
\noindent where the $\cal M$-distribution MGF under LOS blockage is $M_{I}(s|b=0)=M_{I,Kg}^{(1)}(s)$ and the term corresponding to the generalized no blockage case is
\begin{equation}\label{EqMGFMb}
M_{I,b}(s|b=1)=\sum\limits_{k = 1}^{\widetilde{k}}  {\widetilde{m}_k M_{I,Kg}^{(k)}(s)}.
\end{equation}

\section{Outage performance}
Due to the slow-fading characteristic of the free-space optical channel, an appropriate metric for the link performance is the outage probability. Thus, in this section, we derive a novel closed-form general expression for the outage probability in FSO $\cal M$-turbulence links under LOS blockage. In addition, a further asymptotic analysis in the high-SNR regime allows to deduce a simpler approximate expression for the outage probability.
 
\subsection{General analysis}
The outage probability, denoted as $P_{out}$, is defined as the probability that the instantaneous SNR, given by $\gamma$, falls below a specified threshold, denoted as $\gamma_{th}$, which represents a protection value of the SNR above which the quality of the channel is satisfactory. Thus, $P_{out}$ can be expressed as  $P_{out}=\text{Pr}\left\{\gamma<\gamma_{th}\right\}$, where Pr$\{ \cdot \}$ denotes probability, or also in terms of the optical irradiance, using the Eq. \eqref{EqSNR}, as
\begin{equation}
P_{out}=\text{Pr}\left\{I<\sqrt{\frac{\gamma_{th}}{\gamma_0}}\right\}=\text{Pr}\left\{I<\frac{1}{\sqrt{\gamma_n}}\right\},
\end{equation}
\noindent where, as defined in Section 2 (Eq. 2), $\gamma_0$ is the received electrical SNR in absence of turbulence and $\gamma_n=\gamma_0/\gamma_{th}$ is the normalized received electrical SNR in absence of atmospheric turbulence. 

According to the above definition, $P_{out}$ can be obtained from the CDF of the normalized received irradiance under LOS blockage, $F_{I,b}(I)$, since
\begin{equation}\label{Eq_Pout}
P_{out}=\int_0^{\frac{1}{\sqrt{\gamma_n}}}f_{I,b}(I) dI=F_{I,b}\left(\frac{1}{\sqrt{\gamma_n}}\right),
\end{equation}
and, in turn, the $\cal M$-distribution CDF can be defined from Eq. \eqref{Eq_MpdfBloc}, as
\begin{equation}\label{Eq_McdfBloc}
F_{I,b}(I)=P_b F_{I,Kg}(I;\alpha,1,\xi_g)+(1-P_b) \sum\limits_{k = 1}^{\widetilde{k}}  {\widetilde{m}_k F_{I,Kg}(I;\alpha,k,\widetilde{\mu}_k)},
\end{equation}  
\noindent where $F_{I,Kg}(I;\alpha,k,\widetilde{\mu}_k)$ is the CDF associated to the $k$-th Generalized-$K$ sub-channel. This function can be calculated by applying the equivalence between the Meijer-$G$ and the Bessel $K$ functions \cite[Eq. (03.04.26.0009.01)]{Wolfram} to the integral of the Generalized-$K$ PDF defined in Eq. \eqref{EqPDFKg}, leading to the following expression
\begin{equation}
F_{I,Kg}(I;\alpha,k,{\cal I}_k)=\frac{1}{\Gamma(\alpha)\Gamma(k)}(B\cdot I)^{\frac{\alpha+k}{2}}G_{1,3}^{2,1}\left(B\cdot I\left|\begin{array}{c}
1-\frac{\alpha+k}{2} \\
\frac{\alpha -k}{2},-\frac{\alpha-k}{2},-\frac{\alpha+k}{2} \\
\end{array}
\right.\right),
\end{equation}
\noindent where again $B=\alpha k/{\cal I}_k$, $\Gamma(\cdot)$ is the Gamma function and $G_{p,q}^{m,n}(\cdot)$ is the Meijer-G function.

Now, substituting $I=\gamma_n^{-1/2}$ into Eq.\eqref{Eq_McdfBloc}, the outage probability under LOS blockage is derived as
\begin{equation}\label{Eq_PoutBloc2}
P_{out}=P_b F_{I,Kg}\left(\gamma_n^{-1/2};\alpha,1,\xi_g\right) +(1-P_b) \sum\limits_{k = 1}^{\widetilde{k}}  {\widetilde{m}_k F_{I,Kg}\left(\gamma_n^{-1/2};\alpha,k,\widetilde{\mu}_k\right)}.
\end{equation} 
Note that it is also possible to more compactly express $P_{out}$ as the combination of the outage probabilities associated to the signal propagation through each $k$-th Generalized-$K$ sub-channel, $P_{out,Kg}^{(k)}$, governed by the mixture coefficients of the Binomial or Negative Binomial distributions, $\widetilde{m}_k$, as
\begin{equation}\label{Eq_PoutBloc}
P_{out}=P_b P_{out,Kg}^{(1)}+(1-P_b) \sum\limits_{k = 1}^{\widetilde{k}}  {\widetilde{m}_k P_{out,Kg}^{(k)}}.
\end{equation} 

It is worth noting that the derived closed-form expressions given by Eq.\eqref{Eq_PoutBloc2} and Eq.\eqref{Eq_PoutBloc} are general and hold for any value of the SNR. However, taking advantage of the asymptotic behavior of the outage probability in the high-SNR regime \cite{Wang03}, a simpler expression to estimate the performance of a FSO link under LOS blockage can be achieved. To this end, a MGF-based approach, as in \cite{Wang03,Pham15}, is followed below. 

\subsection{Asymptotic analysis}
Assuming the conditions described in \cite{Wang03} and considering that the received electrical SNR,  $\gamma=I^2\gamma_0$, as described in Eq. \eqref{EqSNR}, the outage probability can be approximated for large enough values of $\gamma_0$ by the following expression
\begin{equation}\label{Eq_AsympPout}
P_{out}\approx \frac{a}{D_M}\left(\frac{\gamma_{th}}{\gamma_0}\right)^{D_M/2}=\frac{a}{D_M}\gamma_n^{-D_M/2},
\end{equation}
\noindent where $D_M$ is the diversity order and $a$ is the diversity gain, obtained from the first non-zero derivative of the $\cal M$-distribution PDF at $I=0$,
\begin{equation}
a=\frac{f_I^{(t)}(0)}{t!},
\end{equation}
\noindent where $t=D_M-1$.
When assuming the considerations described in \cite{Wang03}, there is a direct relation between the behavior of the irradiance PDF at $I=0$ and the decaying order of the corresponding MGF at high values of $s$. In this sense, for $s \rightarrow \infty$, the MGF can be expressed as ${M_{I,b}(s)=b_M |s|^{-D_M}+o\left(|s|^{-D_M}\right)}$, with $b_M=a\cdot \Gamma (D_M)$ and $o(x)$ being a polynomial function whose limit verifies that $lim_{x\rightarrow 0}o(x)/x=0$. Then, the MGF verifies also that
\begin{equation}\label{Eq_DM}
\lim_{s \to \infty} s^{D_M} M_{I,b}(s)=b_M=a\cdot \Gamma (D_M).
\end{equation}

In order to obtain the diversity order of the $\cal M$-distribution, let us first calculate the diversity order of the $k$-th Generalized-$K$ sub-channel, denoted as $D_k$, that must verify that $\lim_{s \to \infty} s^{D_k}M_{I,Kg}^{(k)}(s)=b_k$. From Eq. \eqref{EqMGFKg}, and applying that ${_1}F_1(\cdot, \cdot, 0)=1$,
\begin{equation}
\begin{split}
\lim_{s \to \infty} s^{D_k} M_{I,Kg}^{(k)}(s)&=\frac{\Gamma(k-\alpha)}{\Gamma(k)}\left(\frac{\alpha k}{\widetilde{\mu_k}s}\right)^\alpha \lim_{s \to \infty} \frac{s^{D_k}}{s^\alpha}+ \\ & + \frac{\Gamma(\alpha-k)}{\Gamma(\alpha)}\left(\frac{\alpha k}{\widetilde{\mu_k}s}\right)^k \lim_{s \to \infty} \frac{s^{D_k}}{s^k},
\end{split}
\end{equation} 
\noindent in which the first term leads to a finite non-zero value only for $D_k=\alpha$, whereas the second term, for $D_k=k$. So, the diversity order for the $k$-th sub-channel is given by
\begin{equation}
D_k=\min\{\alpha, k\}.
\end{equation}
\noindent The parameter $b_k$ is then
\begin{equation}
b_k=\begin{cases}
\frac{\Gamma(k-\alpha)}{\Gamma(k)}\frac{\alpha k}{\widetilde{\mu_k}} & \text{for } D_k=\alpha \text{  if  } \alpha<k\\
\frac{\Gamma(\alpha-k)}{\Gamma(\alpha)}\frac{\alpha k}{\widetilde{\mu_k}} & \text{for } D_k=k \text{  if  } \alpha>k
\end{cases}.
\end{equation} 

With regard to the $\cal M$-distribution, the diversity order $D_M$ must verify Eq. \eqref{Eq_DM} which, by introducing Eq. \eqref{Eq_MGFM}, leads to
\begin{equation}\label{Eq_DMKg}
b_M=\sum\limits_{k = 1}^{\widetilde{k}}  {\widetilde{m}_k \left[\lim_{s \to \infty} s^{D_M} M_{I,Kg}^{(k)}(s)\right]}.
\end{equation}
\noindent The lowest value of $D_M$ that makes the previous equation be a constant is $D_M=\min\{D_k\}$ for $k=1 \ldots \widetilde{k}$, governed again by the Binomial or Negative Binomial distribution coefficients $\widetilde{m}_k$.

At this point, it must be noted that from the definition of the $\cal M$-distribution parameters and the equivalences described in \cite{OE_MKg}, any combination of $\{\alpha, \beta, \rho\}$ leads to a finite non-zero sequence of $\widetilde{m}_k$ for $k=1 \ldots \widetilde{k}$, except in two particular cases: the first one corresponds to the case of having a blockage event, for which the only non-zero coefficient is $\widetilde{m}_1=1$. The second one is the corresponding to the special scenario with $\rho=1$, i.e. the whole scattered optical power is coupled to the LOS signal component. In this case, as detailed in \cite{OE_MKg}, the coefficients $\widetilde{m}_k=0$ for every $k \neq \beta$ (if $\beta \in \mathbb{N})$ or $k \neq \infty$ (if $\beta \in \mathbb{R})$, turning the $\cal M$-distribution into a Gamma-Gamma distribution. In this situation, our analysis reduces to the case of having a full blockage of the link, which was the scenario analyzed in \cite{Djordjevic16}.

After this discussion and assuming $\rho<1$, since the first-order Generalized-$K$ sub-channel is always active, i.e. $\widetilde{m}_1\neq 0$, the diversity order of the $\cal M$-distribution is $D_M=1$, independently on the turbulence parameters. Thus, Eq. \eqref{Eq_DMKg} can be simplified to only the first term of the summation and, then, the parameter $b_M$ is obtained from
\begin{equation}
b_M=\widetilde{m}_1 b_1=\widetilde{m}_1 \frac{\Gamma(\alpha-1)}{\Gamma(\alpha)}\frac{\alpha}{\widetilde{\mu}_1}=\frac{\widetilde{m}_1}{\widetilde{\mu}_1}\frac{\alpha}{\alpha-1},
\end{equation}
\noindent which can be introduced in Eq. \eqref{Eq_AsympPout} in order to calculate the asymptotic outage probability expression at high SNR values in a $\cal M$-modeled atmospheric optical link, as
\begin{equation}
P_{out}\approx b_M\gamma_n^{-1/2}.
\end{equation}
\begin{figure}[tbp]
 	\centering
 	\includegraphics[width=8cm]{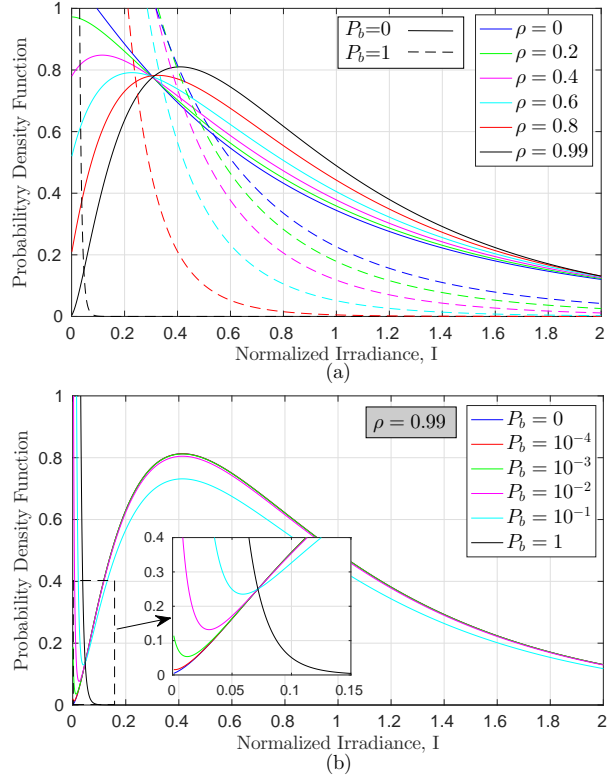}
 	\caption{(a) PDF of the $\cal M$-distribution under LOS blockage for different values of $\rho$ assuming $P_b=0$ and $P_b=1$. (b) PDF of the $\cal M$-distribution under LOS blockage for different values of blockage probability, $P_b$, assuming a very high coupling factor $\rho=0.99$.}
 	\label{FigfIPb0y1}
\end{figure}
Finally, when assuming the partial LOS component blockage with probability $P_b$, as previously described, the outage probability for high SNR  values can be expressed as ${P_{out}=P_b P_{out}(b=0)+(1-P_b)P_{out}(b=1)}$, where $P_{out}(b=1)$ is given by the above equation and $P_{out}(b=0)$, i.e. the outage probability under LOS blockage, is obtained by applying $\widetilde{m}_1=1$ and $\widetilde{\mu}_1=\xi_g$, as shown in Eq. \eqref{Eq_PoutBloc2}.

Hence, the complete outage probability asymptotic expression with LOS blockage is, after some analytical manipulation, 
\begin{equation}\label{Eq_Asymp_PoutB}
P_{out}\approx \frac{\alpha}{\alpha-1} \left[P_b \frac{1}{\xi_g}+(1-P_b)\frac{\widetilde{m}_1}{\widetilde{\mu}_1}\right] \gamma_n^{-1/2}
\end{equation}
\noindent being governed by the first-order Generalized-$K$ sub-channel behavior, in any case.

Inspecting this equation, it is evident that blockage causes the outage probability to be increased for a given value of $\gamma_n$. However, unlike the scenario considered in \cite{Djordjevic16}, the outage probability does not tend to an irreducible floor in the high-SNR regime. This result can be used to design the required power boost to overcome the SNR loss $\Delta$ due to blockage, for a given $P_b$ and a target outage probability as
\begin{equation}\label{Eq_Delta}
\Delta(\text{dB})\triangleq 10\log_{10}\left[\frac{\gamma_n^{(P_b)}}{\gamma_n^{(Pb=0)}}\right]\approx 20\log_{10}\left[1+P_b\left(\frac{\widetilde{m}_1}{\xi_g\widetilde{\mu}_1}-1\right)\right].
\end{equation}

\section{Results and discussion}
In this section, we present results of the effect of partial blockage in both the PDF of $\cal M$-distribution and the outage probability. To obtain these results, we use the closed-form expressions deduced in the previous sections. In all cases, conditions of moderate to strong atmospheric turbulence are assumed with $\alpha=4.2$, $\beta=3$ and $\rho$ ranging from 0 to 1. Note that when $\rho\rightarrow 1$, the amount of optical power coupled to the LOS component is increased, corresponding to a reduction of the turbulence intensity. On the contrary, when $\rho\rightarrow 0$, the amount of optical power coupled to the LOS component is decreased, corresponding to an increase of the turbulence intensity. Thus, by varying the value of $\rho$, turbulence conditions are modified.

First, in Fig. \ref{FigfIPb0y1}, the PDF of the $\cal M$-distribution under LOS blockage, $f_{I,b}(I)$ is shown. In particular, Fig. \ref{FigfIPb0y1}(a) displays the PDF for different values of the coupling parameter $\rho$ (i.e. as a function of the turbulence conditions), while Fig. \ref{FigfIPb0y1}(b) presents the PDF for different values of the blockage probability $P_b$. In both figures, results were obtained using Eq. \eqref{Eq_MpdfBloc} and, so, assuming that an obstacle with diameter $D=D_c$ blocks the coherent component of the optical beam with a probability $P_b$.
\begin{figure}[tbp]
	\centering
	\includegraphics[width=8cm]{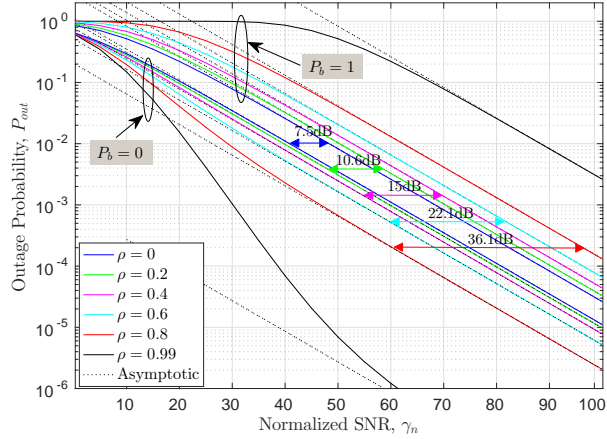}
	\caption{Outage probability of a FSO link over $\cal M$-turbulence channels and LOS blockage with different values of $\rho$ and blockage probabilities $P_b=0$ and $P_b =1$.}
	\label{FigA}
\end{figure}
To cover all possible cases, in Fig. \ref{FigfIPb0y1}(a), two extreme blockage situations are considered. On the one hand, with solid line, the figure shows the PDF of the conventional $\cal M$-distribution without blockage ($P_b=0$) for different values of $\rho$ (i.e. different turbulence conditions). Note that, in this case, when $\rho \rightarrow 1$, the $\cal M$-distribution becomes a Gamma-Gamma distribution. 
On the other hand, with dotted line, the figure shows the PDF of the $\cal M$-distribution with blockage
($P_b=1$). Here, it is clearly observed that as $\rho$ increases the PDF concentrates on smaller values of the normalized irradiance. This is due to a larger portion of the signal being obstructed when turbulence decreases. In fact, in the limit, when $\rho \rightarrow 1$, the distribution degenerates in a deterministic distribution centered at zero. The effect of $P_b$ on the PDF of the $\cal M$-distribution with LOS blockage is shown in Fig. \ref{FigfIPb0y1}(b). Here, we have chosen the worst case of the previous figure (i.e. $\rho=0.99$). It is observed that the behavior of the PDF near $I=0$ is clearly affected as $P_b$ is increased, so that the probability of this range of low irradiances is also increased. Note that these results are in agreement with the scenario described in \cite{Djordjevic16}, since the modified $\cal M$-distribution PDF with $\rho=1$ is given by
\begin{equation}
f_{I,b}(I)=P_b \delta(I)+(1-P_b)f_{I,GG}(I;\alpha, \beta)
\end{equation}
\noindent where $f_{I,GG}(I;\alpha, \beta)$ is the PDF of the Gamma-Gamma distribution with parameters $\{\alpha, \beta \}$. Thus, the blockage probability $P_b$ modulates the relevance among the Gamma-Gamma distributed LOS and the Rayleigh distributed non-LOS components.

Fig. \ref{FigA} shows now the impact of LOS blockage on the outage probability for different values of turbulence conditions (i.e different $\rho$ values).
 To obtain these results, we have used Eqs. (21), (40) and (41).
\begin{figure}[tbp]
	\centering
	\includegraphics[width=8cm]{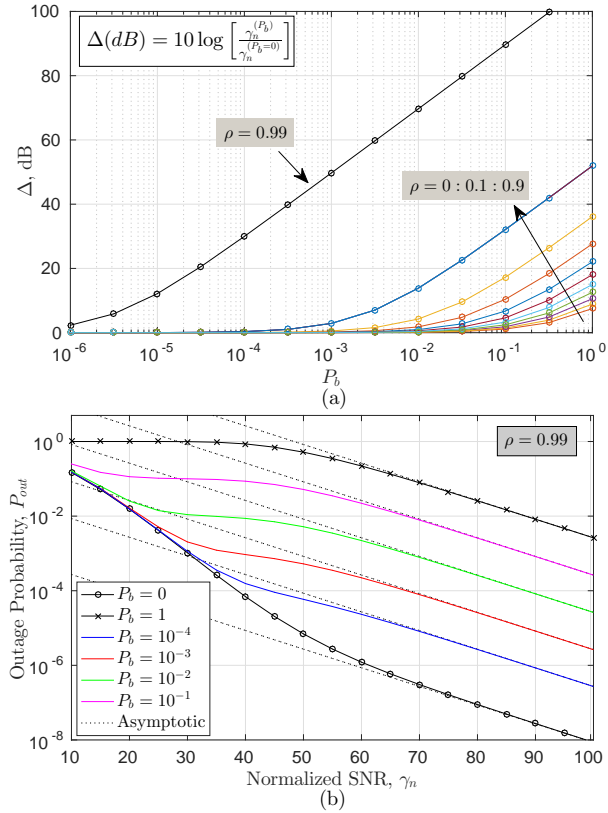}
	\caption{(a) Power boost needed to achieve a $P_{out}=10^{-3}$ under LOS blockage as a function of the coupling factor, $\rho$, and the blockage probability $P_b$. (b) Outage probability of a FSO link over $\cal M$-turbulence channels with $\rho=0.99$ and several values of the LOS blockage probabilities.}
	\label{FigB}
\end{figure}
Again, the extreme situations with $P_b=0$ and $P_b=1$ are first considered. Therefore, the results are grouped in two sets of curves which define the lower and upper bounds of the outage probability.  Note that any blockage probability is between these two bounds. In particular, with solid line are depicted the general results of $P_{out}$ obtained with Eq.\eqref{Eq_PoutBloc2}, while with dotted line are shown the asymptotic results of $P_{out}$ given by Eq.\eqref{Eq_Asymp_PoutB}. Logically, the outage probability is worse in case of LOS blockage ($P_b=1$) than when there is no blockage ($P_b=0$). However, from Fig. \ref{FigA} it is observed that the difference between both bounds decreases when turbulence increases ($\rho\rightarrow 0$). In fact, this difference can be obtained from Eq. \eqref{Eq_Delta} as 
\begin{equation}
\Delta_{max}(\text{dB})=10\log_{10}\left[\frac{\gamma_n^{(P_b=1)}}{\gamma_n^{(P_b=0)}}\right]\approx 20\log_{10}\left[\frac{\widetilde{\mu}_1}{\xi_g\widetilde{m}_1}\right].
\end{equation}
\noindent Note that, for instance, when $\rho=0.8$, $\Delta_{max}=36.1$ dB while for $\rho=0.2$, $\Delta_{max}=7.5$ dB. Thus, these results indicate that the impact of LOS blockage on $P_{out}$ decreases as atmospheric turbulence increases.
The value of $\Delta(\text{dB})$, i.e. the transmit power increment needed to maintain the desired outage probability when a LOS blockage is produced is depicted in Fig. \ref{FigB}(a) as a function of $P_b$ for different turbulence conditions. A target $P_{out}=10^{-3}$ is here assumed. From this figure, it is clearly observed that, since LOS blockage is more harmful as turbulence is reduced (due to more signal being blocked), the value of $\Delta(\text{dB})$ increases as $\rho$ is reduced. So, for $P_b=10^{-1}$ and $\rho=0.1$, a $\Delta\approx 1.4$ dB is needed to maintain the $P_{out}$, while for the same $P_b$ and $\rho=0.9$, $\Delta\approx 32$ dB. Thus, for lower scattered energy coupled to the LOS component, the required power boost is lower and only noticeable for a high $P_b$. On the contrary, higher levels of power increments are necessary to maintain the performance for higher $\rho$ values, even under negligible blockage probabilities.
Fig. \ref{FigB}(b) presents the dependence of $P_b$ on the outage probability. Again, we have chosen the worst case of the previous figure (i.e. $\rho = 0.99$ ). As expected, for fixed turbulence conditions, $P_{out}$ decreases when $P_b$ is reduced. Further, it is also observed that the asymptotic results of $P_{out}$ (depicted with dotted lines) are a good approximation to $P_{out}$ for values of $\gamma_n>50$ dB and $0.2\le \rho \le 0.8$.  Thus, 
Eq.\eqref{Eq_Asymp_PoutB} provides a very simple tool to predict the outage probability behavior of a $\cal M$-modeled FSO link, in the high received SNR regime, with or without LOS blockage.

\begin{figure}[t]
	\centering
	\includegraphics[width=9cm]{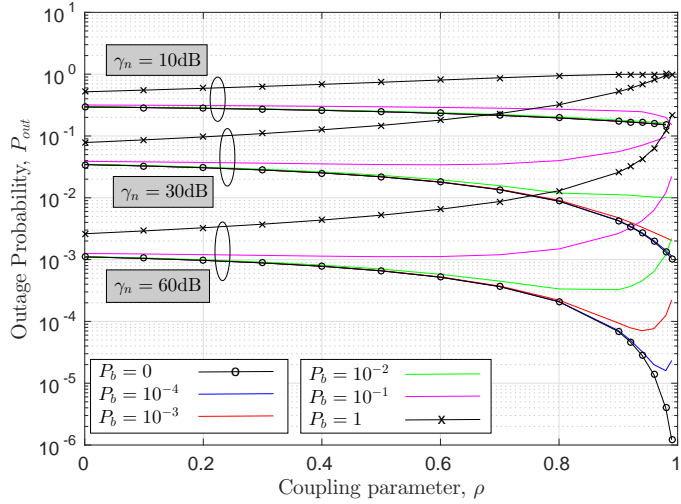}
	\caption{Impact of the coupling factor, $\rho$, on the outage probability for several normalized SNR, $\gamma_n$, and blockage probabilities, $P_b$.}
	\label{FigC}
\end{figure}
Finally, Fig. \ref{FigC} shows the impact of the turbulence conditions, through the coupling parameter $\rho$, on the outage probability considering different values of $P_b$ and $\gamma_n$. 
We observe that for low values of $\rho$ and moderately low values of $P_b$, the outage probability is barely altered. However, as $\rho$ grows, the effect of the LOS blockage becomes prominent. In fact, when $\gamma_n$ is high enough, the value of $P_b$ can change the OP behavior with respect to the turbulence conditions. In particular, when $P_b$ is low ($P_b<10^{-2}$), the OP increases as turbulence conditions worsen ($\rho\rightarrow 0$). However, when $P_b$ is higher ($P_b>10^{-2}$), then, the OP reduces when turbulence conditions worsen. Note that, in this case, an increase in the turbulence intensity can improve the link performance.
In the limit case of $\rho \rightarrow 1$, the outage probability tends to $P_b$; i.e. this scenario reduces to the one assumed in \cite{Djordjevic16}, due to the equivalence between the $\cal M$ and the Gamma-Gamma distributions.

\section{Concluding remarks}
In this paper, we have investigated the effect of partial link blockage on the performance of FSO communication links affected by turbulence fading. We derived new analytical expressions for the PDF and the MGF of the $\cal M$-distribution considering a particular type of partial blockage, referred as LOS blockage, in which only the coherent component of the laser been is obstructed. From these functions, we deduced two closed-form expressions for the outage probability in FSO link under LOS blockage. The first expression is general and valid under any SNR, whereas the second one is an approximation applicable only in the high SNR regime. Since these new expressions can estimate the performance of FSO link affected by both turbulence and blockage, they are valuable tools for FSO system design.
Obtained results demonstrate a strong dependence of the outage performance with the turbulence intensity and with the LOS blockage probability. It is observed that, when LOS blockage is likely, an increase in the turbulence intensity can improve the link performance.



\section*{Funding}
This work was supported by Universidad de Malaga (Plan Propio de Investigacion), and by the Andalucia Talent Hub Program launched by the Andalusian Knowledge Agency, co-funded by the EU Seventh Framework Program, Marie Curie actions (COFUND - Grant Agreement no 291780).
\def\baselinestretch{1}\normalsize

\begin{thebibliography}{99}

  	\bibitem{Kaz10} K. Kazaura, K. Wakamori, M. Matsumoto, T. Higashino, K. Tsukamoto, and S. Komaki, ``RoFSO: a universal platform for convergence of fiber and free-space optical communication networks,'' IEEE Comm. Magazine, {\bf 48}, 130 (2010).
  	
  	\bibitem{Ciaramella} E. Ciaramella et al., ''1.28-Tb/s (32 x 40 Gb/s) free-space optical WDM transmission system,'' IEEE Photon. Technol. Lett., vol. 21, no. 16, pp. 1121--1123, (2009).
  	
  	\bibitem{Bloom} S. Bloom, E. Korevaar, J. Schuster, and H. Willebrand, ''Understanding the performance of free-space optics,'' J. Opt. Netw., vol. 2, no. 6, pp. 178--200, (2003).
  	     
  	\bibitem{And05} L. C. Andrews and R. L. Phillips, \emph{Laser Beam Propagation through Random Media} (SPIE, 2005).

  	\bibitem{Zhu02}
  	X.~Zhu and J.M.~Kahn, ``Free-space optical communication through atmospheric turbulence channels'', IEEE Trans. Commun. \textbf{50}, 1293 (2002).
  	
  	\bibitem{AlH01} M.A. Al-Habash, L.C. Andrews and R.L. Phillips, ``Mathematical model for the irradiance probability density function of a laser beam propagating through turbulent media,'' Opt. Engineering {\bf 40}, 1554 (2001).
  	
  	\bibitem{Churnside} J. H. Churnside, and S. F. Clifford, ``Log-normal Rician probability-density function of optical scintillations in the turbulent atmosphere,'' J. Opt. Soc. Am. A {\bf 4}, 1923--1930 (1987).
  	
  	\bibitem{Toyo11} M. Toyoshima, H. Takenaka, and Y. Takayama, ``Atmospheric turbulence-induced fading channel model for space-to-ground laser communications links,'' Opt. Express {\bf 19}, 
  	15965--15975 (2011).
  		
	\bibitem{CapLibro} A. Jurado-Navas, J.M. Garrido-Balsells, J.F. Paris and A. Puerta-Notario, ``A unifying statistical model for atmospheric optical scintillation'', in ``Numerical simulations of physical and engineering processes'', pp. 181--206, Ed., In-Tech, Sept. 2011 (Invited chapter), ISBN 978-953-307-620-1.
	
	\bibitem{OE_MKg} J.M. Garrido-Balsells, A. Jurado-Navas, J.F. Paris, M. Castillo-Vazquez, and A. Puerta-Notario, "Novel formulation of the $\cal{M}$-model through the Generalized-$K$ distribution for atmospheric optical channels," Opt. Express \textbf{23}, 6345-6358 (2015). 

	\bibitem{Barakat86} R. Barakat, ``Weak-scatterer generalization of the K-density function with application to laser scattering in atmospheric turbulence,'' J. Opt. Soc. Am. A \textbf{3}, 401--409 (1986).
	
	\bibitem{Jakerman87} E. Jakerman and R.J. Tough, ``Generalized-$K$ distribution: a statistical model for weak scattering,'' J. Opt. Soc. Am. A, 1764--1772 (1987).

	\bibitem{Shankar04} P.M. Shankar, ``Error rates in generalized shadowed fading channels,'' Wireless Pers. Commun, \textbf{28}, 233--238 (2004).
	
	\bibitem{Bithas06} P.S. Bithas, N.C. Sagias, P.T. Mathiopoulos, G.K. Karagiannidis, A.A. Rontogiannis, ``On the performance analysis of digital communications over generalized-$K$ fading channels," IEEE Commun. Lett. \textbf{10}, 353--355 (2006).
	
	


	\bibitem{Ghassemlooy2009} Z. Ghassemlooy, W.O. Popoola, V. Ahmadi, and E. Leitgeb. "MIMO free-space optical communication employing subcarrier intensity modulation in atmospheric turbulence channels." In International Conference on Communications Infrastructure. Systems and Applications in Europe, pp. 61-73. Springer Berlin Heidelberg, 2009.
	
	\bibitem{Kolka} Z. Kolka, O. Wilfert, D. Biolek, and V. Biolkova, ��Availability model of free-space optical data link,�� Int. J. Microw. Opt. Technol. 1(2), 612--616 (2006).
	
	\bibitem{Djordjevic16} G.T. Djordjevic, M.I. Petkovic, M. Spasic, and D.S. Antic, "Outage capacity of FSO link with pointing errors and link blockage," Opt. Express \textbf{24}, 219-230 (2016).
	    	
	\bibitem{Hranilovic}
	 S. Hranilovic, ``Wireless optical communication systems,'' (Springer, 2005).
	 
	 \bibitem{Majumdar} A. K. Majumdar and J. C. Ricklin, Free-Space Laser Communica-tions: Principles and Advances. New York, NY, USA: Springer-Verlag, Dec. 2010.
	 	 
	 
	\bibitem{Lerner70}R. M. Lerner and A. E. Holland, "The optical scatter channel," in Proceedings of the IEEE, vol. 58, no. 10, pp. 1547-1563, Oct. 1970.
	
	\bibitem{Abbas05} T. Abbas, K. Sjoberg, J. Karedal, and F. Tufvesson, "A Measurement Based Shadow Fading Model for Vehicle-to-Vehicle Network Simulations, " in Int. J. of Antennas and Propagation, vol. 2015, Art. ID 190607.
	
	\bibitem{Shapiro77} R.S. Kennedy, and J.H. Shapiro, "Multipath dispersion in low visibility optical communication channels," Research Lab. of Elect., RADC-TR-77-73 MIT, 1977.	
	
	\bibitem{OE_PointM} A. Jurado-Navas, J.M. Garrido-Balsells, J.F. Paris, M. Castillo-V�zquez and A. Puerta-Notario, Impact of pointing errors on the performance of generalized atmospheric optical channels," Opt. Express \textbf{20}, 12550-12562 (2012). 
		
	\bibitem{Ansari2016} I.S. Ansari, F. Yilmaz and M. S. Alouini, "Performance Analysis of Free-Space Optical Links Over Malaga ($\mathcal{M} $) Turbulence Channels With Pointing Errors," in IEEE Transactions on Wireless Communications, vol. 15, no. 1, pp. 91--102, Jan. 2016.
	
	\bibitem{Gra00} I.S. Gradshteyn and I.M. Ryzhik, \emph{Table of Integrals, Series and Products} (Elsevier, 2007).

	\bibitem{Wolfram} Wolfram. (2001), The Wolfram functions site, https://functions.wolfram.com
	
	
	\bibitem{Wang03} Z. Wang and G.B. Giannakis, "A simple and general parameterization quantifying performance in fading channels," in IEEE Transactions on Communications, vol. 51, no. 8, pp. 1389-1398, Aug. 2003.
	
	\bibitem{Pham15} V.P. Thanh, T. Cong-Thang and T.P. Anh, "On the MGF-Based Approximation of the Sum of Independent Gamma-Gamma Random Variables," 2015 IEEE 81st Vehicular Technology Conference (VTC Spring), Glasgow, 2015, pp. 1-5.
	
 
\end{thebibliography}
\end{document}